\documentclass[12pt]{article}
\usepackage[latin1]{inputenc}
\usepackage[T1]{fontenc}
\usepackage{amsmath}
\usepackage{amsfonts}
\usepackage{amssymb}
\usepackage{color}
\usepackage{graphicx}
\usepackage{tikz}
\usepackage{tikz-cd}

\newcommand{\be}{\begin{equation}}
\newcommand{\ee}{\end{equation}}
\newcommand{\ba}{\begin{eqnarray}}
\newcommand{\ea}{\end{eqnarray}}

\newcommand{\bea}{\begin{eqnarray}}
\newcommand{\eea}{\end{eqnarray}}

\newcommand{\beq}{\begin{equation}}
\newcommand{\eeq}{\end{equation}}

\newcommand{\fs}
{i\kern+.01em\hbox{\raise.20ex\hbox{$/$}\kern-.58em$s$}}

\newcommand{\bs} {i\kern-.01em\hbox{\raise.25ex\hbox{$/$}\kern-.52em$b$}}
\newcommand{\qs}{/\kern-.52em s}

\newcommand{\dd}
{\kern.06em\hbox{\raise.25ex\hbox{$/$}\kern-.60em$\partial$}}

\begin{document}

\tikzstyle{bag} = [text width=2em, text centered]
\tikzstyle{bag1} = [text width=5em, text centered]
\tikzstyle{end} = []
\title{Dualities and models in various dimensions}

\author{ E.~F.~Moreno$^a$  and
F.~A.~Schaposnik$^b$\thanks{Also at Comisi\'on de Investigaciones Cient\'\i ficas de Buenos Aires, Argentina.} \\ \vspace{0.2 cm} \\
{\normalsize \it $^a$Department of Physics, Northeastern University}\\ {\normalsize \it Boston, MA 02115,
USA.} \\
{\normalsize \it $^b$\it Departamento de F\'\i sica, Universidad Nacional de La Plata}\\ {\normalsize \it Instituto de F\'\i sica La Plata}\\ {\normalsize\it C.C. 67, 1900 La Plata, Argentina}}

\date{\today}

\maketitle
\begin{abstract}
Working within the path-integral framework we first establish a duality between the partion functions of  two $U(1)$ gauge theories with a theta   term  in $d=4$ space-time dimensions. Then, after a dimensional reduction  to  $d=3$ dimensions  we arrive to the partition function of a $U(1)$ gauge theory coupled to a scalar field with an action that  exhibits a Dirac monopole solution. A subsequent reduction to $d=2$ dimensions leads to the partition function of a theory in which the gauge field decouples from two  scalars which have non-trivial vortex-like solutions. Finally this $d=2$  partition function  can be related to the bosonized version of the two-dimensional QED$_2$ (Schwinger) model.
\end{abstract}

\subsection*{Introduction}
The notion of dualities has been a source of  relevant developments in the context of field theories both in high energy  and condensed matter physics. In  the case of gauge theories, electromagnetic duality, already identified in the absence of sources by Faraday and Maxwell, was at the root of Dirac proposal of the possible existence of magnetic monopoles.

The next step in this  context  was the Montonen-Olive conjecture  \cite{MonOli} regarding the existence of  two "dual equivalent" field formulations of the same theory in which electric (Noether) and magnetic (topological) quantum numbers exchange roles. This duality was then recognized as just one example of the so-called  $S$-duality which plays a central role in  supersymmetric quantum field theories and also in string theories (see for example \cite{AGaume} and references therein).

Concerning condensed matter physics, there has been a growing interest in applications of boson-fermion dualities to the study of the quantum Hall regime, U(1) spin liquids, topological
insulators     and quantum phase transitions (see for example \cite{condensed} and references therein).

Inspired by the duality established in   ref. \!\cite{4}  relating  $d=4$ dimensional Maxwell actions with a topological $\theta$-term,  we shall first rederive such connection, now in the framework of the path-integral formulation of quantum field theory starting from an interpolating partition function which allows to connect two  Maxwell-$\theta$ term partition functions $Z_{M\theta}[e,\theta]$ and $Z_{M{\tilde\theta}}[\tilde e,\tilde\theta)$ related through the  $S$-duality group $SL(2, \mathbb{Z})$. We then proceed to a series of dimensional reductions  from $d=4$ to $d=3$ dimensions and from $d=3$ to $d=2$  discussing the resulting partition functions, field equations and their solutions.

\subsection*{The interpolating partition function approach}
Following the approach developed in refs. \cite{EF}-\cite{FE}, we start by introducing an interpolating partition function $Z_I[e,\theta]$ associated to an action $S_I^{(4)}[A,B,C;e,\theta]$   which  includes  three Abelian gauge fields, one of them playing the role of a  Lagrange multiplier. Within the path-integral approach we shall prove the duality discussed by Chatzistavrakidis {\it et al.}\cite{4} at the level of classical actions, now for the quantum partition functions.

The action $S_I^{(4)}[A,B,C;e,\theta]$ reads
\begin{align}
S_I^{(4)}[A,B,C;e, \theta] &=  \frac1{4 e^2}\int \left( F_{\mu \nu}[A] - \frac{i}{2} \epsilon_{\mu \nu \alpha \beta} F_{\alpha \beta}[B]\right) \left( F_{\mu \nu}[A] - \frac{i}{2} \epsilon_{\mu \nu \alpha \beta} F_{\alpha \beta}[B]\right) d^4x \nonumber \\
&+ i \frac\theta{32\pi^2} \int \left( F_{\mu \nu}[A] - \frac{i}{2} \epsilon_{\mu \nu \alpha \beta} F_{\alpha \beta}[B]\right) \left(  \frac{1}{2} \epsilon_{\mu \nu \alpha \beta} F_{\alpha \beta}[A] - i F_{\mu \nu}[B]\right) d^4 x\nonumber \\
& + i \frac{a}{2} \int  \left( \frac{1}{2} \epsilon_{ \mu \nu \alpha \beta}  F_{\alpha \beta }[A] - i  F_{\mu \nu}[B] \right) F_{\mu \nu}[C] \, d^4x
\label{intter2}
\end{align}
where $F_{\mu \nu} [A] =\partial_\mu A_\nu - \partial_\nu A_\mu$  and  $a$ is up to now, an arbitrary constant.

The  interpolating partition function associated to action \eqref{intter2} is then defined as
\be
Z_I^{(4)}[e,\theta] = \int DA DB DC \exp\left({-}S_I^{(4)}[A,B,C;e,\theta]\right)
\label{17}
\ee
Depending on which two fields one integrates out, $Z_I$ will become the partition function for a theory with an action for the remaining one.

Integrating over $C_\mu$ leads to   a delta function ,
$
\delta[(\delta_{\mu\nu}\Box - \partial_\mu\partial_\nu)B_\nu],
$
imposing $B_\mu$ to vanish (up to a pure gauge).  Then, integration over $B_\mu$ becomes trivial and one ends with the partition function for a Maxwell-$\theta$-term theory
\begin{align}\label{cuna}
Z_I^{(4)} &=  \int DA \exp\left\{  -\int  \left( \frac1{4e^2}  F_{\mu \nu}[A] F_{\mu \nu}[A] +i \frac{\theta}{64\pi^2}  \epsilon_{\mu \nu \alpha \beta} F_{\mu \nu}[A] F_{\alpha \beta}[A]\right)
 d^4 x \right\} \nonumber \\
 &\equiv  Z_{M\theta}^{(4)}[e,\theta]
\end{align}

We shall now proceed to obtain a dual action for the field $C_\mu$  by integrating over $A_\mu$ and $B_\mu$. To that end we complete squares in the interpolating action $S_I^{(4)}[A,B,C;e, \theta]$, eq.\eqref{intter2}. In fact, we can write
\begin{align}\label{inserting}
S_I^{(4)}[A,B,C;e,\theta] =& S^{(II)}[A,B,C] + \frac{32 a^2 e^2 \pi^4}{64 \pi^4 + e^4 \theta^2}\int F_{\mu \nu}[C] F_{\mu \nu}[C] \, d^4 x   \nonumber \\
& - i  \frac{4 a^2 e^4 \pi^2 \theta}{64 \pi^4 + e^4 \theta^2}
\int \epsilon_{\mu \nu \alpha \beta} F_{\mu \nu}[C] F_{\alpha \beta}[C] \, d^4 x
\end{align}
where
\begin{align}
S^{(II)}[A,B,C]&= \frac{d_1}{4}\int \epsilon_{\mu \nu \alpha \beta} \left( F_{\mu \nu}[A] - F_{\mu \nu}[B] + u F_{\mu \nu}[C] \right) \left( F_{\alpha \beta}[A] - F_{\alpha \beta}[B] + u F_{\alpha \beta}[C] \right) d^4x \nonumber\\
&+  \frac{d_2}{2}\int \left( F_{\mu \nu}[A] - F_{\mu \nu}[B] + u F_{\mu \nu}[C] \right) \left( F_{\mu \nu}[A] - F_{\mu \nu}[B] + u F_{\mu \nu}[C] \right) d^4x \nonumber\\
&+ \frac{d_3}{4}\int \epsilon_{\mu \nu \alpha \beta} \left( F_{\mu \nu}[A] + F_{\mu \nu}[B] + v F_{\mu \nu}[C] \right) \left( F_{\alpha \beta}[A] + F_{\alpha \beta}[B] + v F_{\alpha \beta}[C] \right) d^4x \nonumber\\
&+  \frac{d_4}{2}\int \left( F_{\mu \nu}[A] + F_{\mu \nu}[B] + v F_{\mu \nu}[C] \right) \left( F_{\mu \nu}[A] + F_{\mu \nu}[B] + v F_{\mu \nu}[C] \right) d^4x
\end{align}
with
\begin{align}
&d_1 = d_2 = \frac{1}{4 e^2} + i \frac{\theta}{32 \pi^2}\; , && d_3 = -d_4 = - d_1^* \nonumber\\
&u = \frac{i a}{4} \frac{1}{d_1}\;,  &&   v= -u^*
\end{align}
Now, changing variables   in the interpolating partition function \eqref{17},
\be
A_\mu \to A'_\mu = A_\mu - B_\mu  + u C_\mu \;, \hspace{1 cm} B'_\mu = B_\mu + A_\mu + v C_\mu
\ee
  $S_I^{(4)}[A,B,C;e,\theta]$ becomes
\begin{align}\label{inserting}
S_I^{(4)}[A,B,C;e,\theta] =& S^{(II)}[A',B'] + \frac{32 a^2 e^2 \pi^4}{64 \pi^4 + e^4 \theta^2}\int F_{\mu \nu}[C] F_{\mu \nu}[C] \, d^4 x   \nonumber \\
& - i  \frac{4 a^2 e^4 \pi^2 \theta}{64 \pi^4 + e^4 \theta^2}
\int \epsilon_{\mu \nu \alpha \beta} F_{\mu \nu}[C] F_{\alpha \beta}[C] \, d^4 x
\end{align}
The terms in  $S^{(II)}[A',B']$ are completely decoupled from $C$ and integration  over $A'_\mu$ and $B'_\mu$  just gives an irrelevant constant ${\cal N}$ so that one ends with
\begin{align}
Z_I^{(4)}[e,\theta]= {\cal N}  \int DC \exp &\left\{ - \frac{32 a^2 e^2 \pi^4}{64 \pi^4 + e^4 \theta^2}\int F_{\mu \nu}[C] F_{\mu \nu}[C] \, d^4 x  \right.  \nonumber \\
& \left.  \quad + i  \frac{4 a^2 e^4 \pi^2 \theta}{64 \pi^4 + e^4 \theta^2}
\int \epsilon_{\mu \nu \alpha \beta} F_{\mu \nu}[C] F_{\alpha \beta}[C] \, d^4 x
 \right\}
\label{cunna}
\end{align}
Now, choosing    $a= {1}/{4\pi}$  we get
\begin{align}
Z_I^{(4)}[e,\theta] & = \int DC \exp \left\{- \frac1{4{\tilde e}^2} \int F_{\mu \nu}[C] F_{\mu \nu}[C] \, d^4 x
+  i \frac{\tilde e \tilde\theta}{64\pi^2} \epsilon_{\mu \nu \alpha \beta} F_{\mu \nu}[C] F_{\alpha \beta}[C] \, d^4 x \right\} \nonumber \\
& = Z_{M \theta}^{(4)}[\tilde e,\tilde \theta]
\end{align}
with
\be
 {\tilde e}^2 = \frac{64 \pi^4 + e^4 \theta^2}{4 e^2 \pi^2} \;, \hspace{1 cm}
{\tilde \theta} = - \frac{4 \pi^2 e^4}{64 \pi^4 + e^4 \theta^2}
\label{chuci}
\ee
Then,  in view of eq.\eqref{cuna}
one has
\be
  Z_{M \theta}^{(4)}[e,  \theta]  =   Z_{M \theta}^{(4)}[\tilde e,\tilde \theta]
  \label{primeraid}
  \ee
One can see that, working at the level of the  interpolating partition function for a theory with action $S_I$  introduced in \eqref{intter2}, we have established a duality between two models with parameters   $(e,\theta)$ and $(\tilde e,\tilde \theta)$ related by eq. \eqref{chuci}. Such result was found  in ref.\cite{4} by relating  the classical   actions. Finally, after a Wick rotation to 4-d Minkowski space,  defining as usual
\be
\tau = \frac{\theta}{2\pi} + i\frac{4\pi}{e^2}
\ee
and using   relations   \eqref{chuci}   one gets   the standard duality  $\bar \tau = - 1/\tau$  which, together with $\theta$ periodicity generates  the $SL(2 , \mathbb{Z})$  group.

\subsection*{Dimensional reductions}

As it is well known, Yang-Mills self-dual instanton equations  in  Euclidean space become, after dimensional reduction, the first order $d=3$   BPS monopole equations  when the $A_4$ gauge field is identified with the Higgs field  and time dependence  is wiped out from all fields \cite{Cervero}-\cite{Manton}.    In this way, the  instanton solution \cite{BPST}  can be connected with the 't Hooft-Polyakov monopole \cite{HP}.  Also,  second dimensional reduction to $d=2$ dimensions can be seen to led to the first order vortex equations of the Abelian Higgs model \cite{Bogomolny}-\cite{deVS}.
We shall here proceed to a series of dimensional reductions of the model we discussed above with the idea of finding solutions of the reduced field equations and also discuss the resulting partition functions.

We now start  to discuss dimensional reductions of the models discussed above. Let us consider the Maxwell-theta term  action $S_{M\theta}^{(4)}$ associated to the partition function $Z_{M\theta}^{(4)}$ defined in eq.\eqref{cuna}
\be
S_{M\theta}^{(4)}[A;e,\theta] =\int d^4x\left( \frac1{4e^2} F_{\mu\nu}[A] F_{\mu\nu}[A] + i \frac{\theta}{64\pi^2} \varepsilon_{\mu\nu\alpha\beta} F_{\mu\nu}[A] F_{\alpha\beta}[A]\right)
\label{lagi}
\ee
with $\mu = \{1,2,3,4\}$.
In order to dimensional reduce the Lagrangian from $d=4$ to $d=3$ dimensions one proceeds as follows: (i) the field dependence is restricted to spatial coordinates $x_i$, ($\{i=1,2,3\}$),   and (ii) the $A_4$ gauge field component is identified with a scalar field $\phi$,
\bea
A_i(x_j;x_4) & \rightarrow & A_i(x_j)    ~~~~~ i,j=1,2,3\nonumber\\
A_4(x_j;x_4) & \rightarrow & \phi(x_j) \label{reduction}
\eea
With this the (Euclidean) interpolating partition function of the resulting reduced theory is given by
\be
 Z_I^{(3)}[e,\theta]  = \exp \int DA_i D\phi \exp(-{S}_{M\theta}^{(3)}[A,\phi; e, \theta])
 \label{PF3}
\ee
with
\be
 { S}_{M\theta}^{(3)}[A,\phi; e, \theta] = \int d^3x \left( \frac1{2e^2} (\partial_i\phi)^2 + \frac1{4e^2} F_{ij}[A ] F_{ij}[A ]  - i \frac{\theta}{16 \pi^2}\partial_i\phi  \varepsilon_{ijk}F_{jk}\right)
 \label{red3}
\ee
Following \cite{Lohe}, we can find solutions to this euclidean 3d model as {\sl static solutions of 4d Minkowski model}. Notice that this requires to absorb the imaginary unit ``$i$'' in the constant $\theta$ (as the Minkowski version of the $\theta$-term does not contain $i$). We will do something more general considering the $\theta$ parameter  a complex constant and find complex field solutions. Since the equations are linear, we can always isolate the real part at the end, if necessary. Therefore, from now on, we will write the action as
\be
 { S}_{M\theta}^{(3)}[A,\phi; e, \theta] = \int d^3x \left( \frac1{2e^2} (\partial_i\phi)^2 + \frac1{4e^2} F_{ij}[A ] F_{ij}[A ]  -  \frac{\theta}{16 \pi^2}\partial_i\phi  \varepsilon_{ijk}F_{jk}\right)
 \label{red4}
\ee
with $\theta$ complex.

As we shall see, we will find a Dirac monopole solution  of the resulting partition function which can be seen, following Polyakov's idea to analyze  confinement  in compact $QED_3$, as instantons in the Euclidean theory \cite{Polyakov}.

In order to see that Dirac  monopoles can also arise as classical solutions of the field equations of action $ { S}_{M\theta}^{(3)}$, we start by integrating  by parts the  last term in ${S}_{M\theta}^{(3)}$. One has
\be
-\frac{\theta}{16 \pi^2} \int d^3x \partial_i\phi  \varepsilon_{ijk}F_{jk} =  \frac{\theta}{8 \pi^2} \int d^3x \phi \partial_iB_i.
\ee
where   $B_i = \frac{1}{2} \varepsilon_{ijk}F_{jk}$
and the surface term vanishes.  Then, for the case of  Dirac monopole configurations, this term in the action cannot be  neglected and  the resulting field equation for $\phi$ takes the form
\be
\nabla^2 \phi =  {\frac{e^2 \theta}{8 \pi^2} }\partial_iB_i
\label{nabla}
\ee
The magnetic field  $B_i$ of a monopole including the Dirac string along the $z$ direction reads
\be
\vec B(\vec x)   = \frac{g}{r^2} \check{r} - g\Theta(-z)\delta(x)\delta(y)\check{k}
\label{magnetico}
\ee
so that the field equation for $\phi$ takes the form
\be
\nabla^2\phi =  \frac{e^2 \theta g}{8 \pi^2} \delta^{(3)}(\vec r)
\ee
where we have used  $eg = 2n\pi$.  From this result  we see that  $\phi$ is given by
\be
\phi_n(\vec x) = - \frac{ne\theta}{16\pi^2}  \frac1{r}
\label{fifi}
\ee
Since the Dirac monopole  satisfies the Amp\`ere law (there is no electric current in the model) and  $\varepsilon_{ijk} \partial_i\partial_j (1/r)=0$, eqs.\,\eqref{magnetico}  and \eqref{fifi} for the magnetic and scalar fields provide a consistent solution for  the system with action \eqref{red3}.  Note that in contrast with the BPS monopole solution in which the scalar has asymptotically a hedgehog behavior, in the present case $\phi$ behaves as a Coulomb potential with a charge  $ne\theta/(4\pi)$.

\noindent The field equations are linear, so we can also have multi-monopole configurations, with monopole located at positions $\vec x_i$,
\begin{align}\label{}
B_\text{multi-m} (\vec x) = \sum_i B(\vec x - \vec x_i) \,
, \quad \phi_\text{multi-m} (\vec x) = \sum_i \vec \phi_{n_i}( \vec x - \vec x_i)
\end{align}

We now proceed to a second dimensional reduction of the partition function $Z_I^{(3)}$  from the $d=3$ to $d=2$ dimensions. In this case we shall identify $A_3$ with a second scalar $\psi$ and all fields will depend just on $x_a$, $a= 1,2$. Now, before this identification it will be convenient to fix the gauge in a way such that the resulting scalar $\psi$ is massive. To this end we shall consider  a gauge fixing \`a la 't Hooft-Feynman inserting in the path integral   \eqref{red4}  the condition
\be
\exp\left(-\frac{1}{2}\mu^2 \int d^3x A_3^2
\right)
\ee
with $\mu$ a parameter with dimensions $[\mu]$ = 1 since in $d=2$ dimensions the fields $A_\mu, \phi$ and $\psi$ should be dimensionless,  $[A_\mu]= [\phi] = [\psi] = 0$.
Note that   limit $\mu^2 \to \infty$ corresponds to fixing the gauge to $A_3 =0$ while the ``Feynman gauge'' can be obtained for $\mu^2 =2$.

We are now ready to identify  $A_3$ with a scalar field $\psi$ and dimensionally reduce   partition function \eqref{red4}. The resulting $d=2$ partition function    $Z_I^{(2)}$ takes the form
\bea
Z_{I}^{(2)}[e,\theta] &=& \int DA_a  D\phi  D\psi \exp \left( - {\cal S}_{M\theta}^{(2)}[A,\phi,\psi; e,\theta] \right)
\eea
where
\begin{align}
{\cal S}_{M\theta}^{(2)}[A,\phi,\psi; e,\theta] = & \int d^2x \left(\frac1{2} (\partial_a\phi)^2 +  \frac1{2} (\partial_a\psi)^2 + \frac{\mu^2}{2} \psi^2 +  \frac1{4} F_{ab}[A ] F_{ab}[A ] \right. \nonumber\\
& \quad \left. -\frac{e^2 \theta}{8 \pi^2}\partial_a\phi  \varepsilon_{ab}\partial_b\psi\right)
\label{red2}
\end{align}
Since the gauge field $A_i$ decouples from the scalars  its field equations read
\begin{equation}\label{maxwell-eqs}
  \partial_a F_{ab} = 0
\end{equation}
while the scalar fields satisfy the coupled equations
\begin{align}
&\nabla^2\phi = \frac{e^2 \theta}{8\pi^2} \varepsilon_{ab}\partial_a\partial_b \psi \nonumber \\
&\left( \nabla^2 - \mu^2\right) \psi = -\frac{e^2 \theta}{8\pi^2} \varepsilon_{ab}\partial_a\partial_b \phi \label{redux}
\end{align}
As in the  $d = 3$ dimensions, we can find classical solutions, in this case scalar global vortex-like  solutions to these equations in terms of Green's functions of the operators $\nabla^2 - \mu^2$, and $\varepsilon_{ab}\partial_a\partial_b$,
\begin{align}
%
\left( \nabla^2 - \mu^2 \right) G_{r, \mu}(x) = 2\pi\delta^{(2)} (\vec x) \,,  \quad \varepsilon_{ab}\partial_a\partial_b G_\phi(\vec x)  = 2\pi\delta^{(2)}(\vec x)
\end{align}
where
\begin{align}
%
& G_{r, \mu}(\vec x ) = - K_0 (\mu r)\\
& G_\varphi(\vec x) = \arctan (y/x)
\end{align}
with $r = \sqrt{(x_1^2 + x_2^2)}$ and $\varphi$ the polar angle.
In terms of these Green's functions a solution to \eqref{redux} can be written  as
\begin{align} \label{vortex}
\phi & = A \, G_\varphi(\vec x)\nonumber\\
\psi & = -\frac{e^2\theta}{8\pi^2} A \, G_{r, \mu} (\vec x)
\end{align}
with $A$ an arbitrary constant. Again, due to the linearity of the field equations, we can consider global multi-vortex  configurations
\begin{align} \label{multi-vortex}
\phi & = \sum_i A_i G_\varphi(\vec x - \vec x_i)\nonumber\\
\psi & = -\frac{e^2\theta}{8\pi^2} \sum_i A_i   G_{r, \mu} (\vec x - \vec x_i)
\end{align}
As a side note, if $\mu = 0$ the vortex-like solutions have a richer structure,
\begin{align} \label{vortex-2}
\phi & = A \, G_\varphi(\vec x) + \frac{e^2\theta}{8\pi^2} B \ln (r) \nonumber\\
\psi & = -\frac{e^2\theta}{8\pi^2} A \, G_{r, \mu} (\vec x) + B \, G_\varphi(\vec x)
\end{align}
with $B$ another arbitrary constant.

We will show that the model described by action \eqref{red2} is dual, in the large $\mu$ limit, to a 2$d$ massive vector field.  In what follows, all the operations are assumed to be done to the partition function ${\cal Z} = \int D\,\text{fields} \; e^{-\mathcal{S}}$, however, for conciseness, we will only record the changes in the action.

The action \eqref{red2} can be written as
\begin{align}
{\cal S} = & \int d^2x \left(\frac1{2} (\partial_a \phi)^2 +  \frac1{2} (\partial_a \psi)^2 + \frac{\mu^2}{2} \psi^2 +  \frac1{4} F_{ab}[A ] F_{ab}[A]  - \frac{e^2 \theta}{8 \pi^2} B_a \partial_a \phi \right) \nonumber\\
& \quad  + i \int d^2 x \lambda_a \left( B_a - \epsilon_{ab} \partial_b \psi \right)
\label{red5}
\end{align}
where $\lambda_a$ is a vector (Lagrange-multiplier) field enforcing the condition
\begin{equation}\label{delta-f}
 B_a = \epsilon_{ab} \partial_b \psi
\end{equation}
We can then  re-write action \eqref{red5} as
\begin{align}
{\cal S} = & \int d^2x \left(\frac1{2} (\partial_a \phi)^2 + \frac1{2} \psi^2 \left( - \nabla^2  + \mu^2 \right) \psi  - i \epsilon_{ab} \partial_a \lambda_b \psi +
 i B_a  \left( \lambda_a + i \frac{e^2 \theta}{8 \pi^2}\partial_a \phi \right) \right) + \mathcal{S}_A
\label{red6}
\end{align}
where $\mathcal{S}_A =  \frac1{4} \int d^2x\, F_{ab}[A ] F_{ab}[A]$.
We shall  now integrate the field $\psi$ in the partition function with action ${\cal S}$, leading to the following  effective action
\begin{align}
{\cal S} = & \int d^2x \left(\frac1{2} (\partial_a \phi)^2 +  \frac1{2} \left(\epsilon_{ab} \partial_a \lambda_b \right) \left( - \nabla^2  + \mu^2 \right)^{-1} \left(\epsilon_{ab} \partial_a \lambda_b \right) + i B_a \left( \lambda_a + i \frac{e^2 \theta}{8 \pi^2}\partial_a \phi \right) \right) + \mathcal{S}_A
\label{red7}
\end{align}
In the large $\mu$ limit, we have
\begin{equation}\label{largmu}
\left( - \nabla^2  + \mu^2 \right)^{-1} = \frac{1}{\mu^2} + \frac{1}{\mu^4} \nabla^2 + \cdots
\end{equation}
Keeping the leading order, we get
\begin{align}
{\cal S} = & \int d^2x \left(\frac1{2} (\partial_a\phi)^2 +  \frac1{4\mu^2} F_{ab}[\lambda]^2 + i  B_a \left( \lambda_a + i \frac{e^2 \theta}{8 \pi^2}\partial_a \phi \right) \right) + \mathcal{S}_A
\label{red8}
\end{align}
where we have used that, in two dimensions,
\[
\left(\epsilon_{ab} \partial_a \lambda_b \right)^2  = \frac{1}{2} \left(\partial_a \lambda_b - \partial_b \lambda_a \right)^2 = \frac{1}{2} F_{ab}[\lambda]^2
\]
Finally, we integrate the field $B_i$ which enforces the condition
\[
\partial \phi_a = - i\frac{8 \pi^2}{e^2 \theta} \lambda_a
\]
so we get
\begin{align}
{\cal S} = & \int d^2x \left(\frac1{4\mu^2} F_{ab}[\lambda]^2 -  \frac{1}{2} \left( \frac{8 \pi^2}{e^2 \theta} \right) \lambda_a^2 \right) + \mathcal{S}_A
\label{red9}
\end{align}
which correspond to a massive vector field $\lambda_a$ (together with a regular gauge field  $A_a$).

We see that action ${\cal S}$ can be identified with a Proca action for a massive (spin 1) vector field in $d=2$ dimensions which is precisely the bosonized version of the QED$_2$  Schwinger model in which the 2-d fermion $\Psi$ with electric charge $e_{SM}$ is coupled to a gauge field which after bosonization acquires a mass $m$ such that $m^2 = e_{SM}^2/\pi$ thus closing the series of dimensional reduction models that we have presented ending with a fermion-gauge field model.

\section*{Summary and Discussion}
Working within the path-integral framework we have established a series of dualities at the level of the   partition functions. We have also  proceed to a series of dimensional reductions from $d=4$ dimensions to $d=3$ and then to $d=2$ and discussed the solutions of the classical field equations of the resulting theories.

Starting from  the interpolating partition function   $Z_I^{(4)}[A,B,C;\theta]$ in $d=4$ dimensions introduced in eq.\eqref{17} and alternatively integrating over $B_\mu,C_\mu$ or $B_\mu,A_\mu$   we
proved the duality between partition functions, $Z_{M \theta}^{(4)}[A;  e, \theta]$ and  $Z_{M \tilde \theta}^{(4)}[C; \tilde e,  \tilde\theta]$ with parameters $(e,\theta)$ and $\tilde e,\tilde\theta)$, related according to eq.\eqref{chuci} which can be seen   corresponds to the standard duality  $\bar \tau = - 1/\tau$ duality so that this, together with $\theta$ periodicity generates  the $SL(2;\mathbb{Z})$  modular group.

We then proceeded to a dimensional reduction  from $d=4 \to d=3$ and then from $d=3$ to $d=2$  obtaining   actions of  bosonic models   founding the  classical solutions of their field equations.  In the former case we found a Dirac monopole solution  for the gauge field and a $1/r$ (Coulomb potential) behavior for the  scalar arising from the $A_4= \phi$ identification, to be compared  with the   BPS solution for the non-Abelian case which corresponds to a 't Hooft-Polyakov monopole with a hedgehog-like scalar.
Concerning the  $d=3 \to d=2$ reduction we ended with a  $A_a$ gauge field ($a=1,2$) and two scalars $\phi$ and $\psi$ with a partition function with action \eqref{red7}. We also solved the associated field equations   finding non-trivial scalar solutions which correspond to global vortex-like solutions. Moreover, by integrating the scalars fields we ended with a Proca action which,  via bosonization, can be finally connected between the bosonic model and QED$_2$.

The   results described  above are summarized in the following graph:
{\small
\begin{center}
\begin{tikzcd}[cramped, sep=normal]
& Z_{M \theta}^{(4)}[A;  e,  \theta] \\
Z_{M \theta}^{(4)}[A, B, C;  e,  \theta]  \arrow[ru, "\int dBdC"]  \arrow[rd, "\int dBdA" ']
\arrow[rr , shorten=10mm , "4 \to 3"] && \mkern-36mu Z_{M\theta}^{(3)}[A,\phi;\theta,e] \arrow[r,"3 \to 2"] &
Z^{(2)}_{M\theta}[A,\theta,e] \arrow[r,"\to \psi"] & Z_{QED_2}^{(2)}[\Psi,A;e_{SM}]\\
& Z^{(4)}_{M\tilde\theta}[C;\tilde e,\tilde \theta]
\end{tikzcd}
\end{center}
}

We expect to discuss applications of the dualities that we have discussed here to problems in quantum field theory as well as in condensed matter.

\vspace{1.2 cm}

\noindent{\bf{Acknowledgments:}}
We would like to thank Carlos N\'u\~nez for helpful comments and suggestions. F.A.S. is financially supported by PIP-CONICET (grant PIP688) and UNLP  grants.

\end{document}